%% file: main.tex
\documentclass[format=acmsmall, review=false, screen=true]{acmart}

\usepackage{booktabs} 

\usepackage[ruled]{algorithm2e} 

\SetAlFnt{\small}
\SetAlCapFnt{\small}
\SetAlCapNameFnt{\small}
\SetAlCapHSkip{0pt}
\IncMargin{-\parindent}

\definecolor{mypurple}{RGB}{119, 69, 198}

\newcommand{\knm}[1]{\textcolor{black}{ #1}}

\newcommand{\cut}[1]{}

\acmDOI{0000001.0000001}

\received{January 2020}
\received[revised]{June 2020}
\received[accepted]{July 2020}

\begin{document}
\title[Social App Accessibility for Deaf Signers]{Social App Accessibility for Deaf Signers}

\author{Kelly Mack}
\orcid{1234-5678-9012-3456}
\affiliation{%
  \institution{Snap Inc. and University of Washington}
  \country{USA}}
\email{danielle.bragg@microsoft.com}
\author{Danielle Bragg}
\affiliation{%
  \institution{Microsoft Research}
  \country{USA}
}
\email{merrie@microsoft.com}
\author{Meredith Ringel Morris}
\affiliation{%
 \institution{Microsoft Research}
 \country{USA}}
\email{maarten@snap.com}
\author{Maarten W. Bos}
\affiliation{%
  \institution{Snap Inc.}
  \country{USA}
}
\email{ialbi@snap.com}
\author{Isabelle Albi}
\affiliation{%
  \institution{Snap Inc.}
  \country{USA}}
\email{amh@snap.com}
\author{Andr\'es Monroy-Hern\'andez}
\affiliation{%
  \institution{Snap Inc.}
  \country{USA}
}

\begin{abstract}
Social media platforms support the sharing of written text, video, and audio. All of these formats may be inaccessible to people who are deaf or hard of hearing (DHH), particularly those who primarily communicate via sign language, people who we call Deaf signers. We study how Deaf signers engage with social platforms, focusing on how they share content and the barriers they face. We employ a mixed-methods approach involving seven in-depth interviews and a survey of a larger population (n = 60). We find that Deaf signers share the most in written English, despite their desire to share in sign language. We further identify key areas of difficulty in consuming content (e.g., lack of captions for spoken content in videos) and producing content (e.g., captioning signed videos, signing into a phone camera) on social media platforms. Our results both provide novel insights into social media use by Deaf signers and reinforce prior findings on DHH communication more generally, while revealing potential ways to make social media platforms more accessible to Deaf signers.

\end{abstract}


%
%


\setcopyright{acmcopyright}
\acmJournal{PACMHCI}
\acmYear{2020} \acmVolume{4} \acmNumber{CSCW2} \acmArticle{125} \acmMonth{10} \acmPrice{15.00}\acmDOI{10.1145/3415196}


\maketitle

\renewcommand{\shortauthors}{Kelly Mack et al.}

\section{Introduction}
Social media platforms are integral to modern entertainment, civic engagement, news dissemination, and interpersonal communication. Because platforms rely heavily on audio and text, accessing content can pose challenges for the 450 million deaf and hard of hearing (DHH) people
worldwide who have significant hearing loss \cite{WHO}. In particular, \textit{Deaf signers} (DHH individuals who use sign language) encounter difficulties sharing and consuming content on social media platforms. Written English can be inaccessible, since American Sign Language (ASL) is a distinct language with its own vocabulary and grammar; \knm{consequently, English literacy of Deaf signers is typically lower than that of their hearing peers (after high school, on average a sixth-grade vs. ninth-grade reading level) \cite{qi2011large, mckee2015assessing}.\footnote{These works focus on ASL and English, though much applies to other signed and spoken languages.} Text-based interfaces also make it difficult to share content in ASL, as sign languages do not have a standard written form.} Another study cites that 30\% of U.S. high school graduates were "functionally illiterate" \cite{marschark2001educating}. Even video-sharing platforms present barriers, for example due to physical difficulties recording two-handed signing while holding or managing a phone camera.    

Prior work has explored basic sharing behaviors and accessibility issues for DHH individuals, some of whom are signers, on the internet (e.g., \cite{conti2017accessibility, maiorana2014technology, shiver2015evaluating}) and social media (e.g., \cite{kovzuh2018challenges, kovzuh2014examining, kovzuh2016content}), but the literature lacks a deeper understanding of sharing behaviors and barriers, particularly on new visual-centric platforms. Past studies tend to limit their focus to Facebook, without examining more recent and popular visual-centric platforms, such as Instagram and Snapchat, which may present unique opportunities and challenges for Deaf signers using a visual language. These prior studies also target DHH users in general, not Deaf signers specifically, whose preference for non-written, movement-based language places unique requirements on communication technologies. For instance, these prior studies found that DHH users share videos less frequently than other media formats, but the studies do not examine the cause. 

In this work, we present two studies with 67 Deaf signers to expand our understanding of the experiences of, and barriers faced by, Deaf signers on social media platforms. We postulate that Deaf signers have unique needs and challenges on platforms due to their preference for an unwritten, visual language, combined with common platforms' reliance on text-based interfaces. Our work deepens the understanding of accessibility needs and barriers in today's popular social media platforms, by both uncovering novel insights and validating and deepening prior work. Our surveys and interviews were guided by the following two research questions:

\textbf{RQ1:} How and with whom are Deaf signers sharing content on social media?

\textbf{RQ2:} What accessibility barriers do Deaf signers face on social media platforms today?

\knm{Our results are novel in the following ways:  1) Unlike prior work, we focus exclusively on \textit{Deaf signers}, which allows us to contextualize findings within Deaf culture and language\footnote{Note that sign language itself is an integral part of Deaf culture.}; our participants described the importance of sign language to their communication and identity, and their desire to share their culture and language with non-signing, hearing individuals. 2) We explore \textit{why} Deaf signers prefer certain methods of communication on social media platforms to others -- often preferring visual methods like ASL but using written English since it's faster, easier, and understood by more people. 3) We explore \textit{who} Deaf signers' audiences are (mainly hearing family members and DHH friends), and how creating content for different audiences may introduce or remove barriers. For example, sharing for hearing and Deaf individuals can be challenging due to language differences. 4) Finally, we investigate how and how many Deaf signers share in ASL on these platforms and the barriers they face in doing so. While between 28\% and 60\% (depending on the context) of our survey respondents share videos with ASL, they face a spectrum of barriers throughout the process, including filming good quality videos of ASL and captioning these videos for hearing audiences.}

\knm{Our results also validate and deepen the following previous findings: 1) We find that uncaptioned content remains a pervasive issue on social media \cite{kovzuh2018challenges, shiver2015evaluating, conti2017accessibility}, particularly due to the large amount of user-generated content. 2) We deepen our understanding of the effects uncaptioned content has on Deaf signers and how they find and consume video content \cite{shiver2015evaluating}. 3) We delve deeper into the discrepancy between the primary mediums that Deaf signers want (sign language) and use (written English) \cite{kovzuh2014examining}, and the accessibility barriers that Deaf signers face (captioning, language barriers, recording challenges) and their effects (missed information, delays, frustration). For example, we provide the first detailed account of the challenges of filming oneself signing, which may discourage sharing signed content.}

\section{Background and Related Work}
We now discuss prior studies of DHH\footnote{We use ``DHH'' (vs. ``Deaf signer'') when describing prior work concerning deaf individuals that does not focus on people who sign.} users' experiences with social media platforms (mostly Facebook). Our work reassesses the social media usage patterns revealed in previous research on a wider range of platforms and provides more depth to understanding how users share (e.g., how, why, and with whom they share).
A full review of studies of other disabled groups' social media use (e.g., visual \cite{morris2016most, gleason2019s, bennett2018teens}, cognitive \cite{burke2010social, reynolds2018m}, and motor \cite{mott2018understanding} disabilities) is beyond our scope. Additionally, we provide background information about sign language in general to provide a better understanding of Deaf signers.

\subsection{Sign Language Background}
Like other sign languages, American Sign Language (ASL) relays information through hand shapes and motions, posture, and facial expressions \cite{bragg2019Sign, NIDCD}. ASL is the primary language of the Deaf\footnote{``deaf'' signifies hearing status and ``Deaf'' signifies cultural identity} community in the U.S. and much of Canada. Moreover, ASL is a source of pride for many Deaf people, and it is an important part of their identity. As one of our interviewees put it: ``Signing is my native language... it's part of my Deaf Identity... I tend to feel [offended] if some of my hearing family won't use signing to talk with me or won't learn ASL.'' ASL is a completely separate language from English, having a distinct grammar and vocabulary \cite{valli2000linguistics}. Consequently, English literacy of Deaf signers is typically lower than that of their hearing peers \cite{qi2011large}. Signed Exact English (SEE) is another method some DHH individuals use, which uses signs (usually from ASL) with English grammar, and therefore is useful to those who learned English before sign language. Sign languages do not have standard written forms, which can create barriers to using text-based platforms.
    
\subsection{Platform-Centered DHH Social Media Usage Studies}
Past studies of DHH people's social media preferences focus on platforms that are primarily text-based. Such studies found Facebook to be one of the most popular platforms in the U.S. for DHH users \cite{kovzuh2014examining, saxena2015social}. However, they do not examine the newer, increasingly popular visual-centric platforms, such as Snapchat and Instagram \cite{PEW}, which may provide a natural avenue for sharing content in sign language.
Because social media platforms offer such a diverse set of features, our study focuses on the features Deaf signers frequently use, rather than on their overall platform use. A previous study that similarly focused on features \cite{kovzuh2014examining} found that,
while a majority of DHH participants (60\%) preferred communicating in sign language \cite{kovzuh2014examining}, sharing videos is the least common behavior \cite{kovzuh2014examining, kovzuh2015enhancing}, indicating a disconnect between wants and behavior, which we examine further. 

\knm{Two studies analyzed the contents of DHH Facebook pages/groups and revealed insight into sharing behaviors on these forums (though not all members had to be DHH) \cite{kovzuh2016content, kovzuh2019utilisation}. In one study, most communities (91.7\%) shared videos, links, and photos, often with accompanying text. In most of these forums (75\%), written text combined with sign language was the most prevalent form of communication, though the study did not provide details on how writing and signing were combined or the percentage of posts that comprised this ``most prevalent'' class \cite{kovzuh2016content}. In the second study, a smaller percent of examined posts contained sign language (19.8\% sign language, 95.5\% written language, with overlapping sets). From most to least popular, non-textual forms of interaction were: the like button, other react buttons, emojis, stickers, ``text delight animations\footnote{E.g., the confetti that pops up when typing ``congratulations on Facebook''}'', and GIFs \cite{kovzuh2019utilisation}. Our work extends these results by investigating sharing behaviors on a broader set of platforms with Deaf signers specifically and is not limited to DHH/sign language related groups and pages. It also adds depth to the findings of these studies by offering insights into who Deaf signers share content with via each sharing method and why they choose to share with a particular method.}

\subsubsection{Text Messaging by DHH Users}
Text messaging is supported by many popular social media platforms (e.g., Facebook, Instagram, Snapchat, and Twitter), and has been found to be highly useful for DHH users \cite{okuyama2013case, power2004everyone, akamatsu2005investigation}. DHH high school students text to make plans or just to chat \cite{okuyama2013case}, and its informal grammar contributes to people favoring text messaging over email \cite{power2004everyone}. Having messaging devices (this paper predated smartphones) allows students to have more freedom, as their parents can reach them at any time \cite{akamatsu2005investigation}.
Students also practiced English more by texting with hearing peers, though this use of the texting device as an assistive technology of sorts for communication perpetuates existing biases in many assistive technologies for DHH people that they should conform to the dominant hearing society and speak/write English, rather than accommodating their preference for sign \cite{gugenheimer2017impact}. Our work reinforces the utility of text messaging for DHH users on social media and investigates the potential language barriers between Deaf signers and users who do not sign.

\subsubsection{Sign Language Video}
Video calls and mobile phones are not fully satisfactory to signers because they typically only show a portion of a user's body and limit the conversation to a 2D space \cite{keating2003american}. To compensate, signers often move or stand up to fit in the camera frame, turning the body so signs can be understood in 2D space, moving closer to the camera for emphasis, and signing slower and with fewer abbreviations \cite{keating2003american}. Additional concerns arise regarding the strain on mobile phone battery and data usage when sharing signing videos on the go \cite{cavender2006mobileasl}.
Video platforms can provide support for filming and sharing signed videos \cite{hooper2007effects, tran2011evaluating, cavender2006mobileasl}. For example, one study adjusted variables that can ease sending a video over a cellular network like video size, frame rate, and compression of non-essential video areas, and measured the effect on viewer comprehension \cite{cavender2006mobileasl}. They found that decreasing frame rate and regionally compressing unimportant areas of the videos provided a better uploading experience while maintaining comprehension.
We investigate how easily Deaf signers can create signed content, and how signing is captured with mobile phones. We also investigate whether signers need or desire a digital form of ASL or for video compression techniques to be integrated into social media platforms.

\subsection{Social \& Communication App Accessibility for DHH users}
Platform accessibility is critical. A 2018 study found a positive correlation between the perceived accessibility of a platform and its use by DHH people \cite{kovzuh2018challenges}.
Still, many communication apps lack features that DHH users value. A 2010 study identified that the most requested app features by DHH mobile phone users included: sound and video streaming, speech-to-text, and text-to-speech \cite{chiu2010essential}. While many of today's apps or phones natively support video streaming and speech-to-text, a follow-up study \cite{alnfiai2017social} found that commonly used apps were missing several of the key features desired by DHH users, such as supporting the use of sign language, having an image library, and large buttons and fonts, and text-to-speech, but apps ``designed for deaf people'' (Say it with Sign, Skype, Facetime, Glide, Visual Hear, and Let Me Hear) included those features more frequently. Providing these features could make social apps more accessible to DHH users. 

\knm{To help combat accessibility issues, one study compiled a list of guidelines for developers \cite{schefer2018supporting}, based on two prior case studies with deaf users. These guidelines state how to make mobile social networking apps more accessible for deaf users, including making buttons highly readable, prioritizing what a user needs to do next, and providing non-auditory feedback. We provide guidelines for social media platform developers based on a larger survey of users, to help enhance platform accessibility for Deaf signers by specifically supporting the sharing of sign language related content.}

Besides app features, social media platforms have a high volume of user-generated content which provides further accessibility issues for DHH users. Lack of high-quality captions and poor audio quality (which is particularly important to hard of hearing users) are common barriers experienced by DHH people in videos both on social media and online in general \cite{conti2017accessibility, maiorana2014technology, kovzuh2018challenges}. 
\knm{ A study examining captioning of content on DHH-related Facebook groups/pages found that a total of 11 out of 59 videos had captions, and none of the user-generated videos on the forums had captions \cite{kovzuh2016content}. Another looked at captions on Deaf-focused Facebook groups, which would likely have a higher occurrence of captioned content, and found that only 18.6\% of videos were captioned \cite{kovzuh2019utilisation}. The effects of uncaptioned content led to frustration and workarounds like searching for text equivalents of videos or even reaching out to the creators of the video to ask for captions \cite{shiver2015evaluating, conti2017accessibility}. Interviews from one study revealed participants thought automatic captions were not of high enough quality, though some believed they were a promising start.}

Our study investigates these known accessibility barriers on today's social media platforms to assess progress made towards addressing them. At the same time, our work answers novel questions related to accessibility to deepen existing findings. For instance, uncaptioned content exists, but how does that affect users outside of frustration, and how do they find video content today? \knm{If they want to search for signed content, how do they do so, since platforms likely do not incorporate sign language detection algorithms \cite{shipman2017speed}.} Prior work showed that users want apps that support sign language \cite{alnfiai2017social}; what specific accessibility barriers exist to sharing sign language on social apps today, particularly in mobile contexts? Our work attempts to answer such questions.

\section{Methods}
\label{sec:methods}

Our study employed a mixed-methods approach with formative interviews that informed the design of a survey.

\subsection{Formative Interviews}
We performed semi-structured interviews guided by a script (included in the Supplementary Materials), but allowing for unscripted questions about details raised by participants. 
The interviews focused on five main areas: background/demographics, communication preferences (signing, written English, photos), platform preferences (photo/video sharing platforms, text sharing platforms), accessibility issues faced while using technology in general, and Deaf culture. We developed the interview questions based on our research goals, prior work in the space, and our own experiences with Deaf signers. 

\begin{figure}
    \includegraphics[width=.9\linewidth]{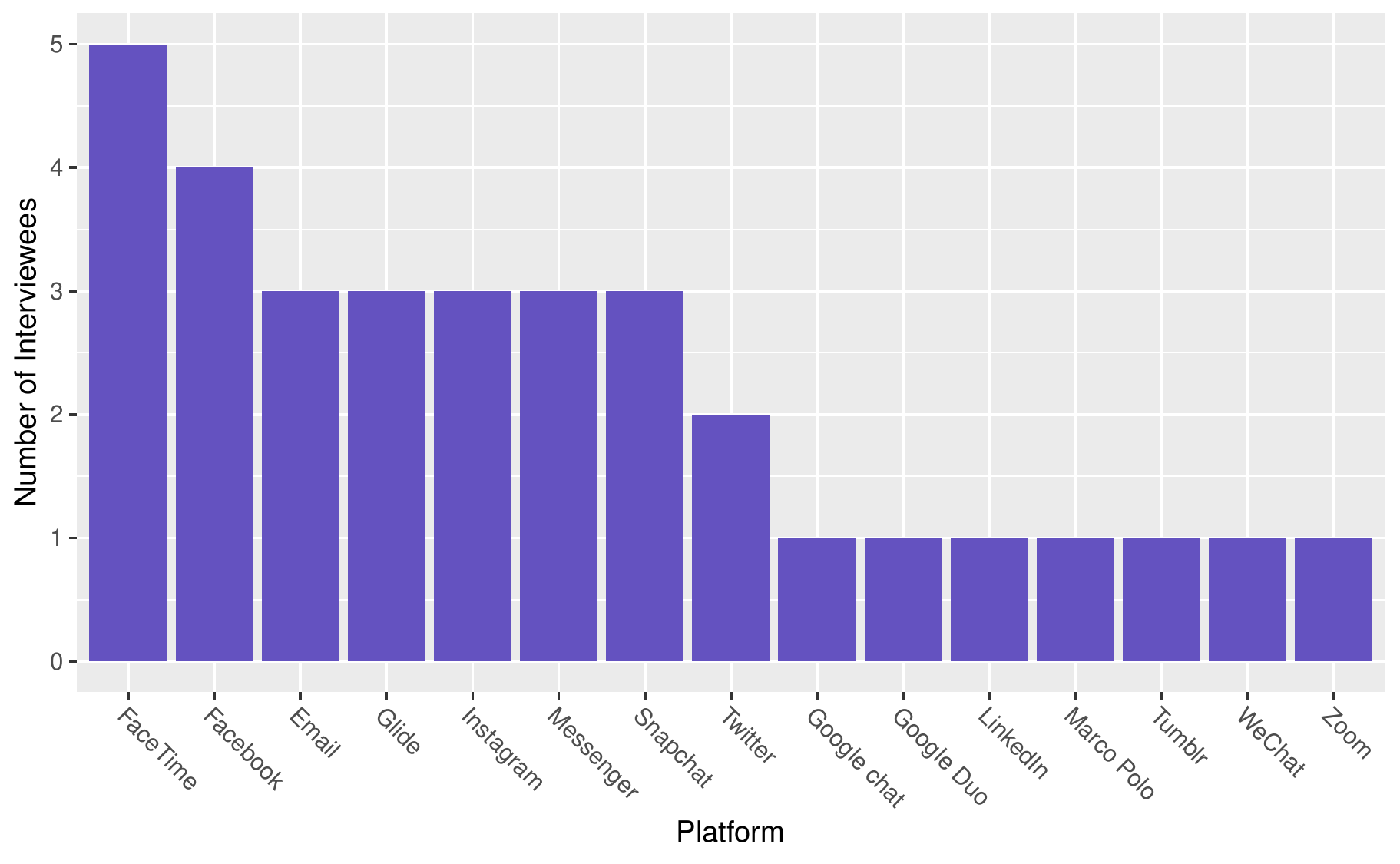}
    \vspace{-1em}
    \caption{Interviewees' reported usage of media types/platforms.}
    \label{fig:platforms}
    \vspace{-1em}
\end{figure}

We recruited seven participants from a list of DHH individuals who had signed up to participate in DHH-related research at a U.S. university. Participants were required to be at least eighteen years old, U.S.-based, use ASL or Signed Exact English (SEE), and identify as DHH. 
Six of our interviewees identified their hearing level as deaf or profoundly deaf, and one stated they were ``deaf and hard of hearing'' (I2).  Table \ref{interview_demo} shows the demographic information of these participants and Figure \ref{fig:platforms} shows the communication platforms they use.

Interviews lasted about forty-five minutes and were conducted via Google Hangouts videochat. Although the interviewer, who knew basic ASL, and the participant could see each other, their conversation occurred via typed messages and did not rely on the video. \knm{Looking back, we realize we should have offered ASL interpretation as an option for participants. This is a limitation of our study. However, as a result, our procedure was more inclusive to users without access to reliable, high speed internet connections, as conducting interviews in ASL requires such a connection. Additionally, we expect our participants to have basic English fluency, as the average reading level of Deaf signers is about a sixth grade level \cite{mckee2015assessing} and the majority of high school graduates are functionally literate \cite{marschark2001educating}, and social media usage requires basic comfort with English (e.g., actualize a profile, navigate menus).} Participants were encouraged to ask questions if anything was unclear. Participants received a \$50 gift card at the end of the interview.

We analyzed the interview transcripts using open coding to identify the main themes in the responses \cite{emerson2011writing, corbin2008basics}. Two researchers read the responses and developed themes and a code book. One of these researchers and a third researcher, who had never seen the data before, then iteratively coded all of the responses according to the code book until they reached agreement (Cohen's Kappa $\geq$ .6 for each theme), discussing all disagreements.

\begin{table}
  \begin{tabular}{llll}
    \toprule
    Interviewee &Gender&Age&Preferred Language\\
    \toprule
    I1&F&25&ASL\\
    \midrule
    I2&F&25&English\\
    \midrule
	I3&F&33&ASL\\
    \midrule
	I4&F&42&ASL/SEE\\
    \midrule
	I5&M&54&ASL\\
    \midrule
	I6&M&45&ASL\\
    \midrule
	I7&M&37&ASL and English equally\\
    \bottomrule
\end{tabular}
  \caption{Interviewee demographics. All participants knew ASL. Note that SEE is not considered a language. Still, one interviewee stated their preferred language was SEE and ASL equally.}
   \label{interview_demo}
\end{table}

\subsection{Themes from Formative Interviews}
After applying the open-coding methods described above, four themes emerged from the interview responses that formed the basis of the subsequent survey instrument. 

\subsubsection{Theme 1: Missing Captions}
\knm{Unprompted, most participants described how frequently they encountered uncaptioned content, one specific, novel scenario being live-streamed videos. Participants stated they felt frustrated by uncaptioned content, which is in line with prior work (e.g., \cite{conti2017accessibility, maiorana2014technology, kovzuh2018challenges}). However, expanding on prior work, our participants' described novel effects of uncaptioned content, like being late to learn about news and pop culture and feeling left out; they also described a variety of novel ways to consume uncaptioned content, like asking other platform users for clarifications. When content did have captions, participants discussed issues with caption accuracy (correct words) and caption timing (properly synced video and captions), which supports findings by Conti et al. \cite{conti2017accessibility}.}

\subsubsection{Theme 2: Preference for Visual Communication}
Participants expressed a strong preference for visual forms of communication. All mentioned using video calling platforms, with a preference for platforms with higher video quality. Six of the seven participants stated ASL as their preferred language. Sign language relies significantly on facial expressions and, relatedly, four participants enjoyed the ability to see or use facial expressions while communicating. Five participants said they like using emojis, GIFs, and memes to communicate. The fact that the majority of interviewees enjoyed these mediums seems to be in contrast to the findings of a 2019 study on European Facebook communities dedicated to hearing loss \cite{kovzuh2019utilisation}. In that study, GIFs, stickers, and emojis were the least popular forms of non-textual sharing. It is unclear whether these different results are due to cultural differences between Europeans and Americans, differences in platform focus (Facebook vs. social platforms more generally), differences between people signers and non-signers, or our small sample size.

\subsubsection{Theme 3: Sharing Sign Language is Difficult}
Participants found recording understandable sign language challenging, and the need for good lighting and full body views difficult to meet on-the-go. \knm{Some of these recording barriers mimicked those found in video calling on a computer with a webcam \cite{keating2003american} (full body view), while others were novel due to the mobile phone form factor (filming with two hands while on-the-go).} Other challenges included video upload and download. One participant mentioned avoiding using video calling platforms or cameras on social media platforms since they drain her battery and use too much data, \knm{which is supported by a body of work looking at ASL video compression \cite{cavender2006mobileasl,tran2011evaluating}. Two participants discussed difficulties sharing accessible content with both signers and non-signers, a novel theme, as sign language can exclude people who do not sign, while spoken/written English can alienate signers who are uncomfortable with English.}

\subsubsection{Theme 4: Sharing with Hearing Individuals can be Difficult}
\knm{The final, novel theme regarded }issues faced when video calling with hearing individuals. Many video calling platforms do not have live captioning although some, like Skype and Google Hangouts, are starting to integrate this feature \cite{skype2015translator}). Users often have to lipread or use the chat feature while calling hearing individuals, which do not always work well. They discussed how difficult lipreading is in person, but it can be made even more challenging when hearing users don't understand the norms of conversation with DHH individuals, such as being well lit and facing the camera.

\subsection{Survey}

We conducted an online survey to explore the themes from our formative interviews with a larger set of participants. Participation criteria were the same as for the interviews (age 18+, U.S.-based, identifies as DHH, uses sign language). We recruited through relevant email lists, social media pages, and a DHH non-profit. The majority of respondents (50) came from this non-profit. 
We donated \$5 to the non-profit for every participant they referred to us and \$5 to the National Association of the Deaf for all other participants. We used snowball recruiting \cite{sadler2010recruitment}, encouraging people to share the survey. \knm{The survey was offered in English only, a limitation that is discussed in section \ref{sec:limits}}

We received 78 responses (completion rate 78\%). We discarded 17 partial responses and one response by a hearing participant, leaving 60 responses for our analyses. We include two who completed everything but the demographics.
Average age was 35.7 (\textit{SD}=12.4). 
Genders were: male (19, 32\%), female (38 63\%), and other (1, 2\%). Preferred languages were: some form of sign language (44, 76\%),  English (10, 17\%), both English and sign language equally (3, 5\%), and "it depends" (1, 2\%).
The average strength of identifying as culturally Deaf on a scale from 1 (not at all) to 7 (very much so) was 6.0 (\textit{SD}=1.5). 
Table \ref{tab:hearing} summarizes their hearing status.

\begin{table}[h]
  \begin{tabular}{lll}
    \toprule
    Hearing Status&\% Respondents&Cannot hear below\\
    \toprule
    Profound&55\%&95 dB\\
    \midrule
    Severe&19\%&80 dB\\
    \midrule
	Moderate&10\%&50 dB\\
    \midrule
	Mild&9\%&30 dB\\
    \midrule
	Other&7\%&{in-between options}\\
    \bottomrule
\end{tabular}
  \caption{Hearing status of survey respondents (n = 60).}
  \label{tab:hearing}
\end{table}

\subsubsection{Survey questions}

The survey questions (provided in Supplementary Materials) were inspired by our interview themes and by prior literature. The questions comprised three main categories:
\begin{enumerate}

\item \textbf{Demographics:} age, gender, hearing status, and comfort with sign language and written English
    
\item \textbf{Communication on Social Media:} how, with whom, and what respondents share

    \underline{Features:} Our interviews revealed that people use the same platforms in very different ways (e.g., communicate with a friend, post to a feed, etc.). Hence, we asked which cross-platform features they use: sharing text, sharing pictures, and sharing videos. \knm{We specifically probed why they chose their primary method of sharing, a novel contribution that goes beyond prior literature on this topic. Additionally, while a few other studies have looked at feature use \cite{kovzuh2019utilisation, kovzuh2016content, kovzuh2014examining}, we differentiate those sharing methods that involve sign from those that don't (e.g., sharing videos with sign vs. those without, written English versus ASL gloss), which other studies do not.}
    
    \underline{Audience:} Interviewees used different communication methods depending on the setting and audience. For example, one interviewee used GIFs with a small group of friends and family members, but preferred video calling with family. \knm{Our survey dove deeper into these differences than prior studies by asking about communication with different audiences in terms of size, hearing status, and social relationship. While other studies have investigated differences based on hearing status \cite{kovzuh2014examining}, to our knowledge, none have added the dimensions of audience size and relationship.}

    \underline{Deaf culture:} Interviewees consistently talked about different ways they shared their connection with Deaf culture on social media. We compiled a list of these methods and asked our survey respondents which they took part in (see Table \ref{tab:DHH_culture}). This question helps us understand in what ways people express Deaf culture and how platforms can better accommodate this expression, \knm{which has not been explored by prior studies. }An additional, open-ended question asked if there were any other parts of Deaf culture respondents wished they could share more easily.

\item \textbf{Accessibility of Social Media:} captions, filming sign language, and sharing videos containing sign language

    \underline{General accessibility barriers:} We compiled a list of accessibility barriers, some novel and discussed by our interviewees and others supported by past literature, and asked our survey respondents to rank how severely each issue impacted them (see Table~\ref{tab:barriers}).
    
    \underline{Video-specific barriers:} Because our interviewees focused on issues creating and sharing sign language videos, we included a question gauging these barriers (see Table~\ref{tab:video_barriers}), \knm{which is a novel addition to the deaf social media literature.} We also asked users to select how they sign to their phone, including options like ``signing with one hand'' and ``propping the phone up to sign with two hands,'' as this could influence the barriers to creating sign language videos. \knm{While barriers to video calling and signing with someone on the computer are somewhat explored \cite{keating2003american}, signing on mobile phones is  less so.}
    
    \underline{Captions:} Our interviewees highlighted that poor quality and missing captions are a problem, which has not previously been confirmed on social media in the U.S. \knm{Prior work on poor captioning has been done either with other languages (German Sign Language) \cite{kovzuh2018challenges} or platforms (the internet generally as opposed to social media) (e.g., \cite{conti2017accessibility, maiorana2014technology}). We verify that these trends hold for our users on social media in the U.S. Additionally, past work has uncovered the effect of frustration that uncaptioned content has on individuals \cite{shiver2015evaluating, conti2017accessibility}; we attempt to reproduce these findings and also investigate a larger array of effects inspired by our interviewees (see Table~\ref{tab:captions}). Finally, we probe their opinions of caption quality when present, as current caption quality results are from a 2017 study of only news sites \cite{conti2017accessibility}. Given recent advances in Machine Learning, in this work, we investigate the state of captions today \cite{chen2019deep}. Furthermore, we look beyond news platforms and to social media platforms.}
    
    \underline{Desired improvements:} We compiled a list of potential features to improve accessibility barriers on social media, some inferred from interviewee experiences (e.g., automatic captioning), and others suggested by interviewees (e.g., creating more ASL-related graphics), \knm{several of which have not been investigated by prior research (e.g., ASL-related graphics, unlimited recording time for videos, and the ability to record hands-free). We asked participants to rank this list by importance (see Table \ref{tab:features}). }
\end{enumerate}




\section{Findings}
Our results show a conflict between desires and behaviors of Deaf signers. While these individuals enjoy communicating visually, barriers on social platforms prevent them from doing so in a timely manner. These barriers include physical ones like recording sign language with a phone and also cultural ones, such as the language barrier between signers and non-signers. Our participants spoke of desires to educate the hearing community and share their culture, though accessibility barriers on social media make that challenging at this time.

\subsection{Communication on Social Media}
While the majority of our respondents preferred sign language, written English was the most common method of sharing due to speed, convenience, and inclusivity to non-signers. All respondents participated in at least one method of engaging with Deaf culture, but also spoke of wanting to share this culture and community with non-signers and hearing individuals.

\subsubsection{Reliance on Written English; Preference of Visual Media}
\label{sec:sharingpreferences}
\begin{figure}
    \includegraphics[width=.6\linewidth]{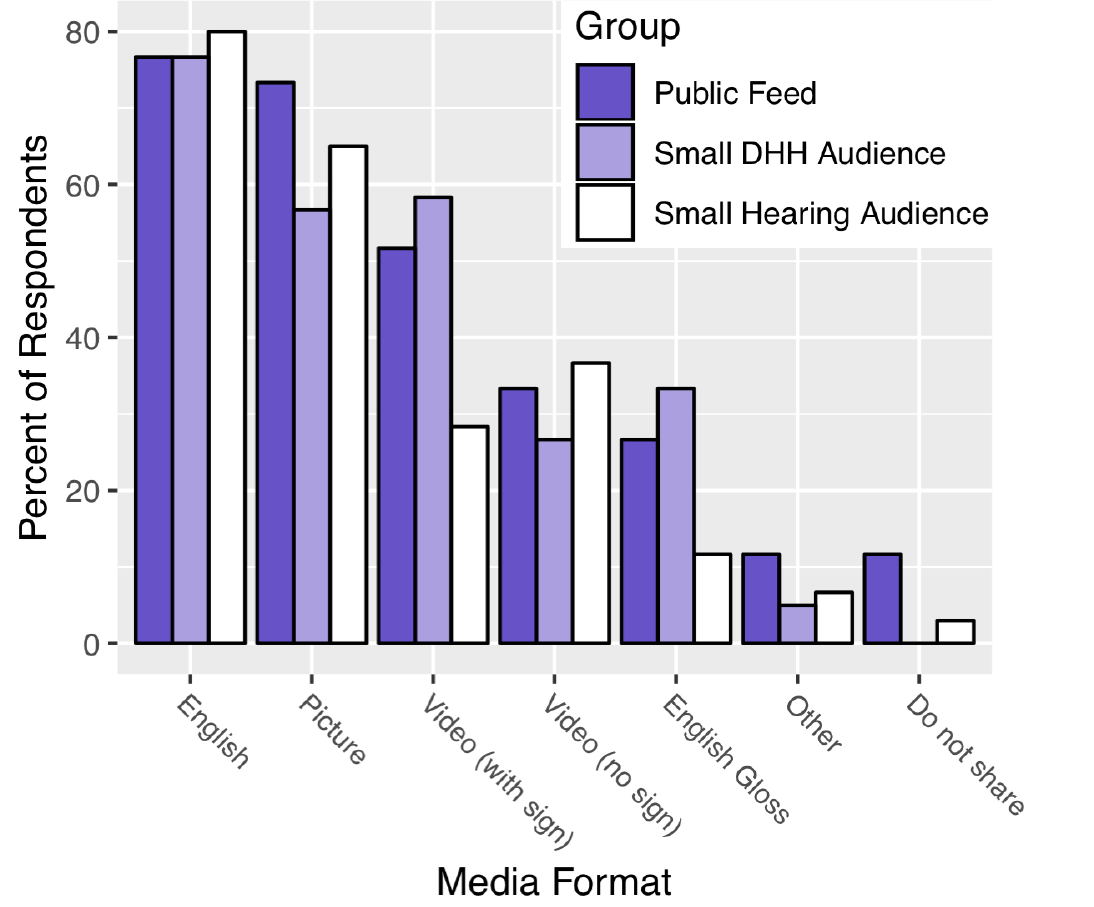}
    \caption{The percentage of respondents who reported sharing with different types of media formats in three settings: 1) a public feed meant for sharing broadly (e.g., a Facebook feed) in dark purple, 2) a small group of DHH individuals (light purple), and 3) a small group of hearing individuals (white). Respondents selected from a list.}
    \label{fig:sharing}
\end{figure}

Figure \ref{fig:sharing} shows what format respondents use when interacting with three types of audiences: public feeds, small DHH groups, and small hearing groups.

Across all audiences, English text was the most popular format for sharing on social media, \knm{as has been shown for DHH individuals in prior studies of Deaf-related Facebook pages and groups in Europe with German Sign Language, but not as a whole on social media or with ASL \cite{kovzuh2016content, kovzuh2019utilisation}. Our survey adds further, novel depth to these observations by exploring \textit{why} written methods are used the most,} finding that written English is used mainly because of its speed, ease of use, and ability to share with all individuals regardless of hearing status.

Even when Deaf signers communicate with groups of DHH individuals, where language barriers would be less likely be an issue, more Deaf signers (almost 20\% more) shared with English than they did with sign language\footnote{Note that many social media platforms share content asynchronously. Non-static ASL could also be shared asynchronously via video clips, GIFs, or animojis}. \knm{This preference mimics the patterns shown with how DHH individuals share with DHH audiences on Deaf-related Facebook forums in Europe \cite{kovzuh2014examining}.}

Among those who selected written English as the format they use the most with public feeds, 80\%  stated they do so for speed and convenience and because it allowed them to communicate with both signing and non-signing individuals. One respondent noted another added benefit of written English: it is more accessible to blind users (because it can be read aloud by screen readers). Speed and convenience were also the most common reason for using written English in small groups of DHH and hearing individuals.

The popularity of English on social media among our participants conflicts with their language preference outside social media: 69\% of respondents preferred some form of signing vs. 18\% who preferred English and 5\% who preferred both English and ASL equally. Interviewees mirrored these preferences: five of seven preferred ASL or SEE\footnote{\knm{Note that SEE is not considered a language, unlike ASL. One interviewee stated their preferred language was SEE and ASL equally.}} and one preferred English and ASL equally. One interviewee explained:
\begin{quote}
    ``Signing is my native language... it's part of my Deaf Identity... I tend to feel [offended] if some of my hearing family won't use signing to talk with me or won't learn ASL'' --\emph{I4}. 
\end{quote}

Respondents share sign language (glossed\footnote{English gloss is written English words following ASL grammar.} or signed) most frequently when communicating with a mainly DHH audience. English gloss and videos containing sign language were most frequently used when sharing with small DHH groups, next-most frequently with public feeds, and least frequently with small hearing groups. Regardless of audience, videos capturing sign language were more popular than English gloss.

Twelve survey respondents stated they shared in ``other'' media formats with at least one of the three audiences. Interviewees gave us insight into these other forms of communication. For instance, interviewees explained loving using graphics like GIFs and emojis due to their (facial) expressiveness: 
\begin{quote}
    ``OH yes [I like sharing GIFs and emojis]... Because we use our facial expressions all the time for ASL'' --\emph{I7}.
\end{quote}

\begin{quote}
    ``I love using GIFs!!! GIFs are [a] way to express my wacky sense of humor to communicate with both D/HH and hearing community'' --\emph{I4}. 
\end{quote}

Participants' value of integrating facial expressions in communication somewhat \knm{contradicts prior research which found GIFs were the least popular method of sharing non-textual data closely followed by stickers and emojis \cite{kovzuh2019utilisation}. This difference could be due to regional, cultural, platform, or signer versus non-signer differences, and should be investigated by further research.}

\subsubsection{Deaf signers share primarily with DHH Friends}

\begin{table}
    \setlength{\tabcolsep}{3pt}
  \begin{tabular}{l|c|c|c|c|c}
    \toprule
    Group&Friends&Family&Coworker&Acquaintance&Other\\
    \toprule
    DHH &90\%&30\%&23\%&40\%&5\%\\
    \midrule
    Hearing &72\%&75\%&40\%&48\%&10\%\\
    \bottomrule
\end{tabular}
  \caption{Survey respondents indicated who they share with when communicating on social media in small groups (DHH and hearing).}
  \label{tab:small_groups}
\end{table}

\knm{While several studies look at strength of Deaf communities on Facebook \cite{kovzuh2016community, kovzuh2016content}, previous studies have not investigated the relationships Deaf signers engage with most on social media.} Our participants used different sharing behaviors with DHH and hearing audiences; their relationships with these audiences also differed (shown in Table \ref{tab:small_groups}). In particular, DHH correspondents were most often friends (reflecting the Deaf community's strength), while hearing correspondents were most often family (as >90\% of DHH children are born to hearing parents \cite{mitchell2004chasing}). Coupled with our earlier results, this breakdown implies that among our participants, most sign language content (e.g. video) is sent between friends, and most content sent within families is English. 

\subsubsection{People express their Deaf culture on social media}
\label{sec:connecting}

\knm{Unlike prior studies focused on Facebook \cite{kovzuh2019utilisation, kovzuh2016content}, we focused on studying how people express their Deaf culture across platforms.} All survey respondents reported at least one way of engaging with Deaf culture on social media, regardless of how strongly they identified as culturally Deaf (see Table \ref{tab:DHH_culture}). The top methods of engaging included belonging to DHH related groups and following/sharing DHH news groups.  

\begin{table}
  \centering
  \begin{tabular}{l|c}
    \toprule
    Method for Connecting with Deaf Community &Percent\\
    \toprule
    I am a part of DHH related groups or pages &68\%\\
    \midrule
     I share/follow DHH news groups &58\%\\
    \midrule
	I share/talk about news related to the DHH community &55\%\\
    \midrule
	I follow DHH celebrities and public figures &48\%\\
    \midrule
	I use ASL/DHH related GIF/meme/emoji/stickers &42\%\\
    \midrule
	I post content which includes English gloss or sign language&42\%\\
    \midrule
	I use jokes, abbreviations, slang, or hashtags related to Deaf culture &38\%\\
    \bottomrule
\end{tabular}
  \caption{How survey respondents expressed their connection with the Deaf community online. The options came from the formative interviews and respondents were asked to select all that apply.}
  \label{tab:DHH_culture}
\end{table}

A free response question, where the participants were not given any specific prompt, revealed respondents wanting to educate hearing people about ASL and Deaf culture. 25\% of respondents reported wanting to share their Deaf culture more easily on social media; five talked about wanting to see more use of and focus on ASL or Deaf events. One respondent asked for more resources for learning ASL, while another wanted to inform hearing individuals about norms for talking with DHH individuals: ``[I wish people knew about the] importance of visual cues to note meaning. Watching people talk with a flat face but hearing their voice change indicates change in meaning but since it's so hard to see it's difficult to tell [if you are DHH].'' Two people talked about raising awareness about the Deaf community, so that ``hearing people [develop] a broader understanding that DHH are not disabled from head to toe!'' Another respondent talked about the spectrum of hearing loss: ``That the Deaf experience is more than just 'hearing' vs 'Deaf'! There is a gray area of deafness that should be more prominently focused on.'' These responses suggest that Deaf signers feel disconnected from (particularly hearing) non-signers. Though theses themes were mentioned by a small number of people, we suggest that future research investigates how to bridge signing and non-signing individuals, since multiple respondents brought up this topic organically.

\subsection{Accessibility Barriers}
At a high level, the top three barriers faced indicated three major areas for concern: lack of captioning, difficulty recording sign language, and language barriers between signers and non-signers. Additional survey questions dived deeper into the themes of lack of captioning and recording sign language. Uncaptioned content led to feelings of frustration and isolation, though caption quality was rated as acceptable on average when present. The strongest, most common video sharing barrier was difficulty in creating captions of signed content for non-signers, again highlighting the language barrier between these groups.
\begin{table}
  \centering
  \begin{tabular}{l|c|c|c}
    \toprule
    Accessibility Barrier&Percent&Mean&SD\\
    \toprule
    The content has no/poor quality captions &77\%&4.6&2.4\\
    \midrule
     It is hard to film good videos (lighting issues, etc.) &70\% & 3.2 & 2.0\\
    \midrule
	I have a hard time communicating with friends who don't know ASL &65\%&3.5&2.4\\
    \midrule
	The apps that I want to use drain my battery &58\%&3.2&2.3\\
    \midrule
	I have a hard time recording myself signing with my phone &55\% & 3.2&2.3\\
    \midrule
	I can't record long enough videos &53\%&2.8&2.1\\
    \midrule
	I can't create content that I can share with DHH \& hearing friends &53\%&2.7&1.9 \\
	\midrule
	I do not have enough data or signal to send/receive the content & 45\%&2.4&2.0\\
	\midrule
	Other & 17\% & NA & NA\\
    \bottomrule
\end{tabular}
  \caption{Surveyed ratings of accessibility barrier severity on a scale of 1 (``this is not an issue for me'') to 7 (``this is a huge issue for me''), where a ranking of 4 indicates that an item was a moderate issue, but any value greater than 1 indicates some level of issue. This table shows the percent facing each (rated>1), mean rating, and standard deviation.}
  \label{tab:barriers}
\end{table}

Participants reported facing a wide range of accessibility barriers (see Table \ref{tab:barriers}). Respondents rated the impact of most barriers as moderately negative, except lack of captioning - the most-often faced barrier - which they rated as having a strong negative outcome. These issues may be more prominent with ASL signers of lower proficiency and older users, as ASL fluency and age were both negatively correlated with difficulty recording sign language with a phone ($\rho$=-.293, \textit{p}<.03 and $\rho$=-.372, \textit{p}<.01, respectively).

\begin{table}
	\centering
    \begin{tabular}{l|l}
        \toprule
        The captions are: & Agreement\\
        \toprule
        Good quality overall & 5.1\\
        \midrule
        Readable & 5.1\\
        \midrule
    	Well timed (speed and synced with video) & 5.1\\
        \midrule
    	Accurate/Error free & 4.2\\
        \bottomrule
    \end{tabular}
    \caption{Survey respondents ranked aspects of caption quality on a scale of 1 (``Strongly Disagree'') to 7 (``Strongly Agree''); 4 indicates a neutral opinion.}
    \label{tab:caption_qual}
\end{table}

\subsection{Captions}
\label{sec:captions}

\knm{Our results confirm that uncaptioned content is prevalent on social media  for our users, and uncover the resulting social impacts on signing Deaf users \cite{shiver2015evaluating, conti2017accessibility}.}
Table \ref{tab:caption_qual} presents respondents' rankings of different aspects of caption quality. Overall, respondents were neutral when rating these aspects, indicating that (when present) caption quality was acceptable; however, the lack of high scores for caption quality indicates that there is still room for improvement. \knm{Nonetheless, these results seem to show improvement over a 2017 study which found many problems with the quality of news site captions which affected experience \cite{conti2017accessibility}; an alternate interpretation is that the caption quality differs between social media and that of news sites or the rest of the web, perhaps due to the quality or content of the videos posted in these different outlets.}

\subsubsection{Uncaptioned content makes users feel lonely and frustrated}

\begin{table}
  \centering
  \begin{tabular}{l|c|c}
    \toprule
    Effects of Uncaptioned Content& Mean &SD\\
    \toprule
    I feel left out &6.1&1.6\\
    \midrule
    I feel frustrated &6.1&1.4\\
    \midrule
	I feel like I miss - or am late to learn about - current news/events &5.5&1.9\\
    \midrule
	I often don't consume the content I like because it does not have captions &5.5&1.9\\
	\midrule
	I often have to spend extra time searching for a version of a video with captions&5.5&1.9\\
    \midrule
	I feel like I cannot relate to pop culture &5.2&2.1\\
    \bottomrule
\end{tabular}
  \caption{Survey respondents indicated how much they agreed with the following statements on a scale of 1 (``strongly disagree'') to 7 (``strongly agree'') about how lack of captioning affects their lives. A response of 4 indicates a neutral opinion}
  \label{tab:captions}
\end{table}

Feeling left out or frustrated were the two most common effects of uncaptioned content on our survey respondents (see Table \ref{tab:captions} for all effects); feelings of frustration were particularly strong for users with stronger Deaf identities (positive correlation of $\rho$=.267, \textit{p}<.05). \knm{The feeling of frustration reproduces prior studies \cite{shiver2015evaluating, conti2017accessibility}, while the feeling of being left out is a new finding.} Similarly, six of our interviewees (86\%) mentioned feeling left out or frustrated when it comes to uncaptioned content, as one interviewee described:
\begin{quote}
``It impacted on my experience to see what is missing [on social media platforms] and what is lacking [compared to the experience of a] hearing audience ... My pleasure of seeing more information through platforms [is] being violated and I want to have the same accessibility as [a] hearing user'' --\emph{I1}.
\end{quote}

Related to feeling left out, the next-most impactful effect was feeling late to learn about news and current events. 
Lack of captioning or interpretation makes content difficult to consume, and many users must wait for an accessible version. 
For example, the growing popularity of live-streaming videos, which often lack captions, was a theme that emerged throughout the interviews. 
Several interviewees elaborated on how this issue affected their news consumption: 
\begin{quote}
``[If] something [is] going on through local news at my hometown ... I had to find out through the share page on [Facebook]. I was too late to find out and very behind to catch up with all details'' --\emph{I1}. 
\end{quote}

\subsubsection{Extra time spent working around uncaptioned content}

Spending time to find accessible versions of uncaptioned content was rated as having a substantial negative impact (5.5 out of 7). Our interviewees elaborated on how to find accessible versions, for example conducting lengthy searches to find captioned versions or similar content, particularly for niche content. 
Another interviewee found that, in an emergency, she could get the information she needed by commenting on the content asking other users online to explain what was happening. \knm{While previous studies found users would ask content providers for captions or transcripts \cite{shiver2015evaluating}, the ability to ask questions to hearing users may be a new affordance social media provides.}

While some people found innovative ways to find consumable content, others did not, blocking them from desired content. 
\begin{quote}
``[I won't look at videos that have no closed captioning.] I look at social media to relax and chill, not [work] hard to understand what the heck [is] going on!'' --\emph{I7}.
\end{quote}
\begin{quote}
``[I watch videos about things I like] not as often as I wish due to lack of [captioning]... if there's no [captions], I do not watch it'' --\emph{I5}.
\end{quote}

\knm{One theme that is underexplored is the correlation between level of comfort with English and effects of uncaptioned content.} We found that the more comfortable a deaf signer was with English and the more hearing they had, the less likely they were to report serious issues with uncaptioned content. For instance, hearing loss was positively correlated with not being able to consume uncaptioned content ($\rho$=.335, \textit{p}<.02), feeling late to learn about news ($\rho$=.313, \textit{p}<.03), and feeling disconnected from pop culture ($\rho$=.405, \textit{p}<.003) -- meaning that someone with more hearing loss is more likely to experience challenges as a result of uncaptioned content. Conversely, English fluency was negatively correlated with taking extra time to find captioned content ($\rho$=-.315, \textit{p}<.02) and feeling late to find out about news ($\rho$=-.317, \textit{p}<.02). These correlations suggest that hearing and comfort with written English lessen the severity of issues associated with uncaptioned content, possibly because higher levels of hearing and comfort with English allow users to find and consume content in other ways than reading the captions.

\subsection{Barriers to Sharing Video with Sign Language}
\label{sec:video_barriers}
\knm{One way our work is most novel is in its exploration of barriers to sharing sign.} With our demographic of Deaf signers, most of whom prefer sign language to English, the ability to share sign is critical, and yet it is still one of the least popular sharing behaviors. In uncovering the barriers to sharing sign, we found some were similar to those barriers experienced while sharing real-time over a computer web camera \cite{keating2003american}, new barriers arose related to capturing sign language on a mobile phone and sharing it with a broad audience on social media.

The two most commonly faced barriers to sharing sign language videos were difficulty in creating captions and the creation/sharing process taking too long (see Table \ref{tab:video_barriers}). We dive deeper into two themes: creating captions and recording sign language with a phone; both can impact the time it takes to share a video. We also investigate the motivation behind sharing preferences for those individuals who most frequently shared in sign language.

\begin{table}
  \centering
  \begin{tabular}{l|c|c|c}
    \toprule
    Video Barrier&Percent&M&SD\\
    \toprule
    It's hard to create captions for my hearing friends &89\%&5.2&2.2\\
    \midrule
    The process of recording \& uploading/sending a video takes too long &77\% & 3.7 & 2.0\\
    \midrule
    Recording video takes too much data to upload/send &72\%&3.8&2.0\\
    \midrule
	Signing into the camera makes it hard for me to sign &72\%&3.7&2.2\\
    \midrule
	It's hard to prop up my phone so I can record myself signing &70\%&3.5&2.1\\
    \midrule
	I don't always look the way I want to in the video &70\% & 3.5&2.2\\
    \midrule
	Recording video drains my battery &68\%&3.5&2.3 \\
	\midrule
	It's hard to find good light to record the video & 65\%&3.1&2.0\\
	\midrule
	I don't like drawing attention to my hearing status & 45\% & 2.3 & 1.9\\
    \bottomrule
\end{tabular}
  \caption{Respondents' rating of accessibility barriers to sharing sign language videos on social media, on a scale of 1 (``this is not an issue for me'') to 7 (``this is a huge issue for me''), where a ranking of 4 indicates that an item was a moderate issue, but any value greater than 1 indicates some level of issue. This table provides the percentage who faced each barrier (severity $>$ 1), mean severity rating, and standard deviation.}
   \label{tab:video_barriers}
   \vspace{-1em}
\end{table}

\subsubsection{Recording challenges: phone constraints, lighting, and space}

Difficulty propping up a phone to record sign language was a common issue ($\sim70\%$ of respondents). ASL has many signs where both hands are doing different things, so ASL is not conducive to being signed with a phone while walking around. A previously proposed solution \knm{to address the challenge of fitting into the frame in a desktop setting was to increase the distance by backing away from the camera \cite{keating2003american}. This could be translated to a mobile phone context by first propping up the phone and then backing up. However, this solution eliminates the ability to sign on-the-go. The alternative is to sign with one hand and hold the phone in the other, eliminating the use of one hand for signing and increasing the difficulty of fitting the body from the waist up in the frame. These trade-offs illustrate the novel problems revealed by exploring the mobile phone as the entry-point for social platforms (vs. prior studies' focus on desktop computers).}

\knm{When asked how they record sign language with their phone, a topic not explored in prior studies,} 58\% of respondents reported propping the phone up against something and signing, which requires users to stop moving and to set up a filming space. 53\% reported signing with one hand and holding the phone with the other. Only 22\% had a phone gadget or case that allows them to sign with both hands while recording. Our findings indicate that filming sign language on-the-go is difficult, unlike texting. The fact that more people take the time to stop what they are doing and prop up the phone can also contribute to the overall lengthy time it takes to record sign language.

People have challenges even when they are stationary and signing into a webcam. If the person is back-lit, it is hard to see their hands and face---poor lighting affected about two-thirds of respondents. If they are too close to the camera, they have a very small space to sign in---about 70\% of respondents stated this was an issue. One interviewee commented on these challenges when video calling with a DHH individual: 
\begin{quote}
``That's usually the first question in our conversations -- can you see me?'' --\emph{I2}.
\end{quote}
These filming challenges can all contribute to the amount of time it takes to create and share a video.

\subsubsection{Adding captions for hearing individuals is time consuming} 
After recording sign language, respondents found sharing the video with their desired audience difficult. The most common issue ($\sim89\%$ of respondents) was difficulty in creating captions so that sign language can be understood by a hearing audience. \knm{To our knowledge, this is the first time this issue has been identified (explicitly raised by Deaf signers in interviews and surveys), though a content analysis of DHH-related Facebook groups and pages found that none of the user-generated videos were captioned (it is unclear if they contained audio or sign language)\cite{kovzuh2016content}.} Two interviewees discussed the time and effort they put into sharing content with both DHH and hearing friends, for example, by adding captions to sign language videos.

\subsubsection{Why people still share with video}
Despite the barriers above, some respondents still preferred sharing sign language videos. Four (7\%) respondents reported using sign language video the most on public feeds because it allowed them to communicate with both DHH and hearing audiences (likely indicating that they captioned or transcribed videos). Three of these four respondents preferred video because it allows them to use their preferred language.

Additionally, twelve respondents reported using videos with sign language the most to communicate with small groups. Eleven of these twelve did so because it allowed them to use their preferred language, seven because it allowed them to be creative when sharing, and only five because signing was fast for them. These findings further support the idea that filming sign language is difficult and time consuming, though it seems the benefits outweighed the drawbacks for these users.

\subsubsection{Desired Accessibility Features}
\label{sec:ally_deisired}

\begin{table}
  \centering
  \begin{tabular}{l|l|l}
    \toprule
    Feature&M&SD\\
    \toprule
    The ability to record videos hands free &6.2&1.5\\
    \midrule
    Automatic captioning & 5.9 & 1.6\\
    \midrule
	ASL graphics (emojis, stickers, GIFs) &5.7&1.8\\
    \midrule
	No time limit for recording videos&5.7&1.7\\
    \midrule
	Video Relay Services (VRS)* & 5.6 &1.9\\
    \midrule
	Text to speech services & 5.2 &2.0\\
    \bottomrule
\end{tabular}
 \caption{Survey prompt, mean, and standard deviation of how important proposed social media platform features are, on a scale of 1 (``not important at all'') to 7 (``very important''),  where a ranking of 4 indicates that an item was moderately important, but any value greater than 1 indicates some level of importance. *VRS supports DHH users with calling hearing people by providing an intermediary ASL interpreter; many social media platforms now allow for real-time video/audio calling between users.}
  \label{tab:features}
\end{table}

Survey participants most wanted accessibility features that improve the recording and sharing of sign language videos, in particular the ability to record hands-free and automatic captioning (Table \ref{tab:features}). Though automatic captioning and hands-free video recording are similarly rated, participants reported a larger negative effect for lack of automatic captioning than for difficulty recording sign language, and more respondents experienced it (77\% vs. 55\% of respondents). It is possible that many respondents simply abandoned recording videos, thereby bypassing captioning difficulties, but would re-engage with better support.

\knm{In these rankings of potential accessibility features, participants show support for all of the options listed in Table \ref{tab:features}, including the novel themes introduced by our interviewees and survey results: creating ASL graphics, ensuring there is no time limit for recording, or hands-free recording. An added benefit of these three features is that social platforms  can implement these with existing technologies.}

One survey respondent requested automatic captioning not just for spoken content, but for ASL: 
\begin{quote}
    ``Create [a] feature that recognizes hand shapes in ASL that will be able to voice what the signing video is about or create auto-subtitles of what we are signing in the video. Our goal is to expose and educate the hearing community about what the Deaf community [is like].''
\end{quote}
This desire for captions of both ASL and English indicates a desire to make ASL content more accessible to non-signers and supports our earlier finding that the most common accessibility barrier with sharing videos was difficulty making video captions. Therefore, captioning of both sign language and spoken content are desired features for our respondents.

\section{Discussion}
While answering our two research questions, we identified many issues that detract from Deaf signers' experience while using social media, such as language barriers, challenges to sharing videos of sign, and captioning issues. Based on these findings and prior work, we subsequently explore these issues in further detail, and discuss how platforms might mitigate these issues. 

\subsection{RQ1 - DHH Communication on Social Apps}
Our results revealed two main themes related to RQ1: 1) a desire of signing Deaf social media users to connect with hearing people and non-signers, and 2) a desire to better integrate sign language into platforms.

\subsubsection{How can social apps connect Deaf signers and hearing people?}

\knm{Our interview and survey findings revealed a desire of Deaf individuals for hearing people to better understand the Deaf community, which was not mentioned in prior studies in a social media context (though this has been studied in general communication scenarios, e.g., \cite{gugenheimer2017impact}).}
The Deaf community has a strong and rich culture, of which many hearing people are unaware. Social media provides a powerful opportunity to connect these two groups and their cultures, as many platforms already have hearing and DHH membership. \knm{We now present novel ways to utilize these platforms to do so.}

\textbf{Remove Language Barriers between Signers and Non-Signers.} 
Reducing the language barrier between Deaf signers and non-signers is important for encouraging communication and understanding between groups. This sentiment is reflected in prior work stating many Deaf signers are not comfortable with English \cite{qi2011large} and our survey respondents (I feel left out- 6.1/7; I feel like I cannot relate to pop culture- 5.2/7; I feel like I miss - or am late to learn about - current news/events- 5.5/7; I have a hard time communicating with friends who don't know ASL- 3.5/7). \knm{Though our study selected for people who had some comfort with English, our sample did include individuals with low literacy in written English (5 individuals ranked their comfort as less than 4 out of 7); we suspect this selection bias underestimates rather than overestimates the need to remove language barriers.}

A potential solution could be for social platforms to employ professional sign language interpreters to interpret videos; while cost-prohibitive at scale, adding captions and/or interpretations to popular content could be a starting point. This solution would also provide valuable training data to aid computer vision research on automated sign language recognition. \knm{Another potential solution is to provide options for lexical simplification in platforms. Though studies have not yet concluded direct benefits for reading comprehension, a 2020 study did find DHH participants preferred to have the autonomy to change the language complexity of text on demand \cite{alonzo2020automatic}.}

\textbf {Share Deaf Culture and Norms.}
Another way to facilitate communication between signers and non-signers is to encourage hearing individuals to learn more about Deaf culture, which spans art, poetry, music, and more \cite{ladd2003understanding,holcomb2012introduction}.  Many social media platforms already create creative tools and content for users to engage with (e.g., Facebook stickers, Snapchat Lenses, Instagram filters). Platforms could create similar creative content tailored to sign language and Deaf culture, specifically around Deaf cultural events like Deaf awareness month (September). Highlighting content by Deaf celebrities who span both Deaf and hearing communities (e.g., Nyle DiMarco) may be another way to facilitate cultural exchange.

Moreover, hearing users can learn norms to facilitate interactions between the groups. For example, outside of social media, during interpreter-facilitated conversations, DHH individuals prefer that hearing people speak to them, not their interpreters \cite{smith2008signing}. Similarly, when conversing over video calls the norm is for hearing individuals to be well-lit and facing the camera. Several of our interviewees explicitly mentioned that hearing users are often unaware of these norms, causing communication difficulties and stress for the DHH user. Platforms could gracefully teach these norms to users and even provide automated feedback on some of these aspects. \knm{These types of solutions support Deaf culture and the use of sign language as opposed to pressuring Deaf individuals to adhere to the majority culture, which is found with many assistive technologies meant to support communication and connection between hearing and deaf individuals \cite{gugenheimer2017impact}.}

\subsubsection{How can social platforms better integrate sign language?}

Social media platforms are typically based in English or another written language, from terms of service to menu content to user-generated posts and messages. This design creates barriers for users with lower literacy rates, including many Deaf signers \cite{mckee2015assessing}. Even among those comfortable with English, many Deaf signers still prefer sign language, as indicated by both our interviewees and survey respondents.

\textbf{Supporting Sign Language Character Systems.}
\knm{Incorporating sign language character systems, textual or animated \cite{bragg2018designing}, that represent signs directly (e.g., HamNoSys, SignWriting, Si5s) may allow ASL users to more fully engage with the entirety of platforms without the need for building separate, video-based features for Deaf signers. 
However, alternate character systems are currently impractical, as none are widely accepted; indeed, focusing on written sign formats may conflict with the need to support (rather than influence) Deaf culture.} 

\textbf{Supporting Signing Avatars}
A more traditional way to integrate sign language in social media is to facilitate consumption through signing avatars. Signing avatars provide signed interpretations of written text on the screen, from user posts to terms of service (see \cite{kennaway2001synthetic,verlinden2001signing, animate2012CUNY, al2018modeling, huenerfauth2009sign}). However, sign language translation and avatar rendering are not fully solved, offering a rich area for research that may enable future solutions, rather than immediate ones. \knm{Researchers could benefit from accounting for demographic and experiential factors, such as type of school attended and technology experience, when exploring any potential solutions \cite{kacorri2015demographic}.}

\subsection{RQ2 - Accessibility of Social Media}
Our interviews and survey revealed two main categories of accessibility issues that Deaf signers face on social media platforms: difficulty in sharing videos with sign language and difficulty in finding captioned content.

\subsubsection{How can platforms support sharing sign language videos?}
While many participants preferred ASL over English, they mostly communicated in written English on social media. Given this apparent contradiction, a plausible explanation is that there are significant barriers and/or inconveniences within the pipeline of creating and sharing a video, which our survey and interview data support (see \ref{sec:video_barriers}). We now discuss how we might improve each stage of the video sharing pipeline: recording, uploading, and consumption.

\textbf{Convenient Recording.} The second and third most frequently faced barriers for our survey respondents were taking too much time to create and upload a video and not having a wide enough field of view on the camera to capture their signing, both affecting about 75\% of respondents. Several physical solutions could make recording sign language easier. A gadget that allows a user to film with two hands while the camera is suspended in front of them would support recording two-handed signing, remove restrictions on mobility, and reduce time spent finding a space to place the camera. Selfie sticks allow users to create more space between the camera and themselves (adding more space in which to sign), but still require one hand to hold the stick. Finding novel form factors to support wide-angle, well-lit, two-handed signing is an area for future design innovation. Another way to expedite recording sign language would be better camera technology that would automatically illuminate the face and hands while darkening the background, thus reducing back-lighting issues.

\textbf{Efficient Uploading.} About half of respondents reported having insufficient battery life or data/signal to upload a video once it is created (58\% and 45\% respectively). Prior work (e.g., \cite{hooper2007effects, tran2011evaluating, cavender2006mobileasl}) investigated video compression to decrease the data load of cellular network uploads. In settings where users send videos back and forth quickly (i.e., videos made to be seen only once, like Snaps on Snapchat), a regional compression algorithm targeting the background while maintaining signing integrity and facial expressions could provide an acceptable trade-off between video quality and upload cost. 

\textbf{Captioning for Hearing Consumption.}  The most common and severe video-sharing barriers faced by our participants was the difficulties in creating captions of signed content (89\% of respondents; 5.2/7 severity). This desire for captioning their own content may be driven by a desire to be inclusive; having experienced the negative effects of inaccessible content, Deaf individuals may want to ensure that they do not do the same to others. For example, one survey respondent noted a preference for written English because it is accessible to blind users. While the need for captioning audio content for DHH individuals is well documented, the need to provided captions of signed content for hearing consumers is novel and in need of attention. 

There seems to be an important distinction between captions and transcripts to Deaf signers, and a desire for captions specifically. A transcript is a static piece of text annotating video or audio content, whereas captions appear and disappear in real-time with the corresponding content. Many social media platforms (e.g., Facebook, Snapchat) allow users to enter text along with a posted video, which could be a transcript. 
Despite this affordance, support for easier caption creation was the most requested feature. \knm{There is currently no consensus in the literature on which is more beneficial (captions or transcripts) \cite{conti2017accessibility, shiver2015evaluating}, and though our work suggests captions may be more favored, deeper exploration of DHH and Deaf signer preferences is a valuable avenue for future investigation.} One solution for creating captions and transcripts is automatic recognition of sign language, but this remains a challenging problem for computer vision, with no widely used solutions \cite{bragg2019Sign} \cite{ye2018recognizing}.

It is currently a painstakingly slow and tedious process to add captions to videos of sign language. While automatic sign language recognition and captioning could help address this problem, these capabilities have not yet been developed, and are areas for future work. As an intermediary step, partially automated process could be used to facilitate manual captioning, for instance, syncing an English transcript with a signed ASL video. Once this technology is developed, it can be incorporated into social media platforms.

\textbf{How can platforms improve captioning of spoken language?} A clear theme supported by prior work, our interviews, and our survey is that lack of captioning in online content is a persistent problem for DHH individuals (see \ref{sec:captions}). Existing guidelines specify that audio content should be captioned \cite{WCAG, ADA, CVA}, but are not well enforced. Additionally, social media sites contain a large amount of user-generated content, which is often uncaptioned and can be difficult to caption, particularly for live-streamed videos.

Improved captions would benefit not only DHH users, but all users. Captions benefit people learning a language (to reinforce the spoken language \cite{markham2001effects, markham1989effects}), and people in loud or quiet places (e.g., airports or libraries). By providing video metadata, captions also improve search engine indexing and discovery. Improved captioning technology could produce all of these benefits.

\textbf {Automatic Captioning of Spoken Language.}
Though automatic captioning exists, it is not accurate enough for all social media content. 
Both Microsoft PowerPoint and Google Slides provide automatic, live captioning of presentations, though this problem is simplified by cues in slide content \cite{googleslides, powerpoint}.
YouTube automatically captions content, though YouTube personalities and Deaf advocates point out unacceptable inaccuracies \cite{youtubecaptionfail, atlantic}. \knm{Indeed, prior work found that DHH individuals did not think automatic captioning was accurate enough on its own \cite{shiver2015evaluating}.} Though imperfect automated captioning technology may have utility in the non-live setting, for example by providing a starting point for human captioners. Additionally, within-platform crowds can be harnessed to correct incomplete, inaccurate, or mis-synchronized captions, for example, via social microvolunteering \cite{brady2015gauging, lasecki2012real}. Social media platforms are particularly well-tailored to these types of tasks, as they have a large number of dedicated daily users that could be leveraged as a part of the crowd and can offer incentives such as titles or badges.

\subsection{Limitations and Future Work}
\label{sec:limits}
\knm{One limitation of our study design is our reliance on English in the interview and survey, which is not the primary language of many Deaf signers, who may prefer ASL. As a result, our sample skewed towards ASL users with relatively high levels of comfort with English. Though, we did still recruit users who were not as comfortable with English (ranked 4 or less on a scale of 1 to 7 where 7 is maximum comfort), they comprised only 8\% of our total sample. An avenue for future work would be to offer a similar survey and interview with ASL interpretation as an option, though this would limit geographic reach and place harsher internet requirements on participants.} 

\knm{To summarize and add to points from our discussion, we suggest the following avenues for future work. A number of themes evolved from our study that merit further investigation. These include facilitating easier recording of sign language on a mobile phone as well as creating ways to facilitate adding captions to signed content to ease the burdens of the video sharing pipeline. Along these lines, clarification as to the benefits and drawbacks of captions versus transcripts in different scenarios, as transcripts are more economical to produce though captions are generally preferred \cite{shiver2015evaluating}, needs further investigation. In addition, the number of participants who organically discussed their desire to educate hearing, non-signers about Deaf culture and Deaf people indicates another avenue for potential, valuable research. Finally, it would be interesting to conduct similar studies to ours with both signers and non-signers and in both English and ASL, to see if the results of this study extend to a broader population. }

\section{Conclusion}
\knm{We investigated how social media platforms can better accommodate Deaf signers by asking 67 individuals about how they share and the barriers they face on popular platforms. We provide depth to current literature on DHH social media use by focusing specifically on Deaf signers, their sharing behaviors on a variety of platforms, who they share with, and why they choose to share with the methods they do. While many Deaf signers prefer to share in sign language, they resort to sharing in English on platforms due to barriers in sharing videos. We discuss in depth these novel barriers including difficulty in creating captions, difficulty filming sign, and language barriers. Based on the insights found in this study, we present many feasible changes for platforms and areas for future research to better support Deaf signers. We hope this work inspires similar in-depth explorations of the Deaf community's unique usage of various interfaces.}

\section{Acknowledgements}
We would like to thank Deaf Planet Soul for helping us recruit participants for our survey, as well as all who participated in the interviews and survey. We would also like to thank Hannah Murphy and for helping with the data analysis.

\appendix
\section*{Appendix}
\input{appendix.tex}
\bibliographystyle{ACM-Reference-Format}
\bibliography{sources}

\end{document}

%% file: appendix.tex
\section{Interview Script}
\label{sec:interview-script}

The following is the script we used as a starting point during our semi-structured interviews. Some of these questions were not asked in all interviews depending on time constraints and applicability.

First, we are going to talk a bit about your background. If you are not comfortable answering any of these questions, please let me know.
\begin{itemize}
    \item What is your age?
    \item How would you describe your hearing status?
    \item What gender do you identify as?
\end{itemize}

We are going to talk about the methods you use to communicate with you friends and family. By methods we mean things like video chatting, writing/typing English, signing, sending pictures, sending emojis, sending gifs, writing sign language, etc. Again, if you are not comfortable answering any of these questions, please let me know.

\begin{itemize}
    \item What is the language you first learned? Is this your preferred language today?
    \item Can you tell me about the methods of communication you use with your DHH friends/family when you are not physically together? 
    \item Can you tell me about the methods of communication you use with your DHH friends/family when you are not physically together? 
    \item What is your favorite method of communication with DHH individuals?
    \item Can you tell me about the methods of communication you use with your hearing friends/family when they are not physically together? 
    \item What is your favorite method of communication with hearing individuals?
    \item Do you use any written form of sign language (including English glossing)?
    \item Are you able to lip read? How is it for you? Do you enjoy it?
\end{itemize}

Now, we are going to begin talking about the different platforms you use to communicate. By platform, we mean any apps, websites, or other technology mediated services you use to communicate (examples include Facebook, Google Hangouts, SMS/text messaging). We are also going to talk about how you share content on social media platforms. We will use the terms broadcasting (sharing with a wide audience, e.g., a Snapchat or Instagram story, a Facebook post to all of your friends) and narrowcasting (creating content to be shared with a few individuals e.g. sending a Snapchat or text message to one or two friends).

\begin{itemize}
    \item What platforms do you use to communicate with hearing individuals?
    \item What platforms do you use to communicate with deaf individuals?
    \item Are there platforms you avoid using to communicate with hearing individuals?
    \item Are there platforms you avoid using to communicate with DHH individuals?
    \item What is your preferred platform for communication with hearing individuals?
    \item What is your preferred platform for communication with DHH individuals?
    \item What type of content do you like to share via broadcasting (videos, pictures, sentences, memes, etc)?
    \item What type of content do you like to share via narrowcasting (videos, pictures, sentences, memes, etc)?
    \item On average, do you think you produce or consume more public content?
    \item Are there any additional functionalities you wish certain platforms supported?
    \item Can you tell me about how the lack of these desired functionalities impacts your experience on these platforms?
    \item Are there any ways you have figured out how to work around these barriers?
    \item How frequently do you find video content that is not closed captioned? Many times a day? About every day? About every week?
    \item If you find videos with CC, can you tell me about the quality of the captions you see?
    \item How does lack of CC affect your experience on these platforms?
    \item Do you ever feel excluded from content that you either found while looking at publicly shared content or received from direct communication? Can you provide an example?
    \item Are there any ways you've figured out how to work around lack of CC?
    \item If you find videos with CC, can you tell me about the quality of the captions you see?
    \item Are there any platforms that you know of or use because they are particularly accessible to DHH users? How did you learn about these platforms?
    \item Are there any platforms that you avoid specifically because they are not accessible? Please explain.
    \item Do you ever share the same content with DHH and hearing individuals?
    \item Do you ever run into problems sharing the same content with both hearing and DHH individuals?
    \item Do you use Snapchat? Why or why not?
    \item Do you record yourself signing ASL/SEE with your phone? Can you tell me about that process?
    \item Do you ever have issues with things like lighting of the video (yours or the other person in the conversation) or with the fact that the camera only captures your upper body?
    \item Do you use gifs, emojis, memes, or similar methods of communication in ways that are specific to just you and your friends/family who know ASL/SEE?
    \item Are there any other parts of Deaf culture (norms, phrases, jokes, traditions) which you find ways to incorporate into your communication on these platforms?
    \item Are there any aspects of Deaf culture which you wish you could express more easily (or at all) on these platforms?
\end{itemize}

\section{Survey Questions}
\label{sec:survey}

The following is a complete list of the questions we asked in our survey.

For these next questions, when we say small group, we mean a group of 10 people or less.
These questions refer to all types of social media apps/platforms, not just those listed as examples.

\begin{enumerate}
\item Which of the following features do you use on any social media platform? (Select all that apply)
\begin{itemize}
	\item Chat/direct messaging with an individual or small group (e.g., Facebook Messenger, Instagram DM, etc.)
	\item Image/video messaging with an individual or small group (e.g., Snapchat, Facebook Messenger, etc.)
	\item Sharing of text or news with a large group (e.g., Facebook feed, Twitter, etc.)
	\item Sharing of images/videos with a large group (e.g., Snapchat stories, Facebook feed, etc.)
	\item None of these
\end{itemize}

\item Select all that apply, while interacting on social media with a large audience (e.g., Facebook feed, Snapchat stories, Twitter, etc.) I:
\begin{itemize}
	\item Share my thoughts in written English (posts, status update)
	\item Share my thoughts in English gloss (English with ASL grammar)
	\item Share my pictures
	\item Share my videos which contain signing
	\item Share my videos which do not involve signing
	\item Re-share content created by other users
	\item React to shared content (liking/commenting on content)
	\item I don't share with a large audience
	\item Other (fill in box below):
\end{itemize}

\item Which method do you use the most to communicate on social media with a large audience?
\begin{itemize}
	\item Share my thoughts in written English (posts, status update)
	\item Share my thoughts in English gloss (English with ASL grammar)
	\item Share my pictures
	\item Share my videos which contain signing
	\item Share my videos which do not involve signing
	\item Re-share content created by other users
	\item React to shared content (liking/commenting on content)
	\item Other (fill in box below):
\end{itemize}

\item Why do you [insert answer from previous question] the most on social media with a large audience? (Select all that apply)
\begin{itemize}
	\item It is fast and convenient
	\item I am able to use my native language
	\item I can be more creative with this feature
	\item I can share content with both hearing and DHH people via this method
	\item It is popular/trendy
	\item Other (fill in box below):
\end{itemize}

\item Select all that apply, while interacting on social media with a individual or small group of people who are DHH (e.g., Facebook Messenger, Snapchat, etc.) I:
\begin{itemize}
	\item Share my thoughts in written English (posts, status update)
	\item Share my thoughts in English gloss (English with ASL grammar)
	\item Share my pictures
	\item Share my videos which include signing
	\item Share my videos which do not include signing
	\item Re-share content created by other users
	\item React to shared content (liking/commenting on content)
	\item I don't share with DHH groups/individuals
	\item Other (fill in box below):
\end{itemize}

\item Which method do you use the most to communicate on social media with an individual or small group of people who are DHH?
\begin{itemize}
	\item Share my thoughts in written English (posts, status update)
	\item Share my thoughts in English gloss (English with ASL grammar)
	\item Share my pictures
	\item Share my videos which include signing
	\item Share my videos which do not include signing
	\item Re-share content created by other users
	\item React to shared content (liking/commenting on content)
	\item Other (fill in box below):
\end{itemize}

\item Why do you [insert answer from previous question] the most on social media with an individual or small group of people who are DHH? (Select all that apply)
\begin{itemize}
	\item It is fast and convenient
	\item I am able to use my preferred language
	\item I can be more creative with this feature
	\item It is popular/trendy
	\item Other (fill in box below):
\end{itemize}

\item Who typically make up the small groups of DHH people you communicate with on social media? (Select all that apply)
\begin{itemize}
	\item Family
	\item Friends
	\item Coworkers
	\item Acquaintances
	\item Other (fill in box below):
\end{itemize}

\item Select all that apply, while interacting on social media with an individual or small group of people who are hearing (e.g., Facebook Messenger, Snapchat, etc.) I:
\begin{itemize}
	\item Share my thoughts in written English (posts, status update)
	\item Share my thoughts in English gloss (English with ASL grammar)
	\item Share my pictures
	\item Share my videos which include signing
	\item Share my videos which do not include signing
	\item Re-share content created by other users
	\item React to shared content (liking/commenting on content)
	\item I don't share with hearing groups/individuals
	\item Other (fill in box below):
\end{itemize}

\item Which method do you use the most to communicate on social media with an individual or small group of people who are hearing?
\begin{itemize}
	\item Share my thoughts in written English (posts, status update)
	\item Share my thoughts in English gloss (English with ASL grammar)
	\item Share my pictures
	\item Share my videos which include signing
	\item Share my videos which do not include signing
	\item Re-share content created by other users
	\item React to shared content (liking/commenting on content)
	\item Other (fill in box below):
\end{itemize}

\item Why do you [insert answer from previous question] the most on social media with an individual or small group of people who are hearing? (Select all that apply)
\begin{itemize}
	\item It is fast and convenient
	\item I am able to use my preferred language
	\item I can be more creative with this feature
	\item It is popular/trendy
	\item Other (fill in box below):
\end{itemize}

\item Who typically make up the small groups of hearing people you communicate with on social media? (Select all that apply)
\begin{itemize}
	\item Family
	\item Friends
	\item Coworkers
	\item Acquaintances
	\item Other (fill in box below):
\end{itemize}

\item How much are you affected by the following issues while using any social media platform? (Ranked each on a scale of 1 [this is not an issue for me] to 7 [severely impacts my experience])
\begin{itemize}
	\item I cannot record long enough videos									
	\item I have a hard time recording myself signing with my phone	
	\item It is hard to film good videos (lighting issues, etc.)
	\item I do not have enough data or signal to send/receive the content	
	\item The apps that I want to use drain my battery									
	\item The content has no/poor quality captions									
	\item I have a hard time communicating with friends who don't know ASL
	\item I cannot create content that I can share with my DHH and hearing friends
\end{itemize}
									
\item If there are any other issues you face, please let us know here.

\item To what extent do you agree with the following statements when you find videos on social media that do not have captions? (Ranked each on a scale of 1 [strongly disagree] to 7 [strongly agree] where 4 is neutral)
\begin{itemize}
	\item I feel left out									
	\item I feel frustrated									
	\item I feel like I cannot relate to pop culture									
	\item I feel like I miss - or am late to learn about - current news/events
	\item I often have to spend extra time searching for a version of a video with captions
	\item I often don't consume the content I like because it does not have captions
\end{itemize}
										
\item When videos do have captions, to what extent do you agree with the following statements about the captions? (Ranked each on a scale of 1 [strongly disagree] to 7 [strongly agree] where 4 is neutral)
\begin{itemize}
	\item The captions are good quality overall									
	\item The captions are readable (good font size and color)									
	\item The captions have good timing (not too fast, synced well with video)
	\item The captions are accurate/error-free
\end{itemize}

\item How important is it that a social media platform have the following? (Ranked each on a scale of 1 [not at all important] to 7 [very important])
\begin{itemize}
	\item Automatic captioning									
	\item Video Relay Services									
	\item ASL graphics (emojis, stickers, GIFs)									
	\item No time limit for recording videos									
	\item The ability to record videos hands free									
	\item Text to speech services
\end{itemize}

\item How do you record yourself signing with your phone's camera (front or back)? (Select all that apply)
\begin{itemize}
	\item I hold my phone in one hand and sign with the other hand
	\item I prop my phone up on something and sign with two hands
	\item I have a case/gadget which helps me prop my phone up and sign with two hands
	\item I do not record myself signing with my phone
	\item Other (fill in box below):
\end{itemize}

\item To what extent do any of the following prevent you from creating and sharing a video of you signing? (Ranked each on a scale of 1 [this is not an isseu for me] to 7 [this is a huge issue for me])
\begin{itemize}
	\item It's hard to find good light to record the video									
	\item It's hard to prop up my phone so I can record myself signing
	\item Signing into the camera which only captures part of my body									
	\item I don't always look the way I want to in the video (i.e. my hair is a mess, I don't want people to see my messy room, etc.)									
	\item It's hard to create captions for my hearing friends									
	\item Recording video drains my battery									
	\item Recording video takes too much data to upload/send									
	\item The process of recording and uploading/sending a video takes too much time	
	\item I don't like drawing attention to my hearing status
\end{itemize}

\item How strongly do you agree with the following statement: I would be more likely to share videos of myself signing if the content was automatically deleted after a short period of time (e.g., after it is opened or after a few hours).
\begin{itemize}
	\item Strongly disagree
	\item Disagree
	\item Neither agree nor disagree
	\item Agree
	\item Strongly agree
\end{itemize}

\item How strongly do you identify as Deaf culturally?
\begin{itemize}
	\item 1 (Not at all)	
	\item 2	
	\item 3	
	\item 4	
	\item 5	
	\item 6	
	\item 7 (Very strongly)
\end{itemize}

\item How do you express your connection with the Deaf community online
\begin{itemize}
	\item I follow DHH celebrities and public figures
	\item I am a part of DHH related groups or pages
	\item I share/follow DHH news groups
	\item I share/talk about news related to the DHH community
	\item I use ASL/DHH related GIFs/memes/emojis/stickers
	\item I post content which includes English glossing or sign language
	\item I use jokes, abbreviations, slang, or hashtags related to Deaf culture (e.g. Champ, true biz)
	\item I do not express/do not have a connection to the Deaf community
\end{itemize}

\item Are there any parts of Deaf culture that you wish you could convey more easily (or at all) on social media?

\item What is your age (years)?

\item What gender do you identify as?
\begin{itemize}
	\item Male
	\item Female
	\item Other/Non-binary
\end{itemize}

\item How would you describe your hearing status?
\begin{itemize}
	\item Hearing
	\item Mild hearing loss (can't hear sounds below 30 dB)
	\item Moderate hearing loss (can't hear sounds below 50 dB)
	\item Severe hearing loss (can't hear sounds below 80 dB)
	\item Profound hearing loss (can't hear sounds below 95 dB)
	\item Other (fill in box below):
\end{itemize}

\item Do you use a cochlear implant or other assistive hearing devices?
\begin{itemize}
	\item No, I do not use assistive hearing devices
	\item Yes, I use hearing aids
	\item Yes, I use a cochlear implant
	\item Yes, I use a different assistive hearing device (fill in box below):
\end{itemize}

\item Do you have any of the following non-hearing related disabilities?
\begin{itemize}
	\item Visual impairment
	\item Cognitive impairment
	\item Motor impairment
	\item No, I don't have any other impairments
	\item I'd rather not say
	\item Yes, I have another disability called (fill in box below):
\end{itemize}

\item What was the first language you learned?
\begin{itemize}
	\item English
	\item ASL
	\item Other (fill in box below):
	\item FavLanguage
	\item What is your favorite  language to use to communicate?
	\item English
	\item ASL
	\item Other (fill in box below):
\end{itemize}

\item How comfortable are you with writing in English?
\begin{itemize}
	\item Extremely comfortable
	\item Moderately comfortable
	\item Slightly comfortable
	\item Neither comfortable nor uncomfortable
	\item Slightly uncomfortable
	\item Moderately uncomfortable
	\item Extremely uncomfortable
	\item I don't use written English
\end{itemize}

\item How comfortable are you with signing ASL?
\begin{itemize}
	\item Extremely comfortable
	\item Moderately comfortable
	\item Slightly comfortable
	\item Neither comfortable nor uncomfortable
	\item Slightly uncomfortable
	\item Moderately uncomfortable
	\item Extremely uncomfortable
	\item I don't use ASL
\end{itemize}

\item Is there anything else you would like to share with us?

\item Did you have any difficulties completing the survey? Were any questions confusing to you?

\item Are you interested in being contacted with follow up questions from this survey or about future research related to the survey your took today? Your email address will only be used for this purpose.
	\begin{itemize}
		\item Yes and my email is
		\item No
	\end{itemize}
\end{enumerate}